\documentclass[graybox]{svmult}

\usepackage{mathptmx}       
\usepackage{helvet}         
\usepackage{courier}        
\usepackage{type1cm}        

\usepackage{makeidx}         
\usepackage{graphicx}        
\usepackage{multicol}        

\makeindex             

\begin{document}

\title*{Contact processes and moment closure on adaptive networks}
\author{Anne-Ly Do and Thilo Gross}
\institute{Anne-Ly Do \at Max-Planck-Institute for the Physics of Complex Systems, N\"othnitzer Str. 38, 01187 Dresden, Germany \email{ly@mpipks-dresden.mpg.de}
\and Thilo Gross \at Max-Planck-Institute for the Physics of Complex Systems, N\"othnitzer Str. 38, 01187 Dresden, Germany \email{gross@physics.org}}

\maketitle

\abstract{Contact processes describe the transmission of distinct properties of nodes via the links of a network. They provide a simple framework for many phenomena, such as epidemic spreading and opinion formation. Combining contact processes with rules for topological evolution yields an adaptive network in which the states of the nodes can interact dynamically with the topological degrees of freedom. By moment-closure approximation it is possible to derive low-dimensional systems of ordinary differential equations that describe the dynamics of the adaptive network on a coarse-grained level. In this chapter we discuss the approximation technique itself as well as its applications to adaptive networks. Thus, it can serve both as a tutorial as well as a review of recent results.}

\section{Introduction}
\label{cpsec:1}
Contact processes are based on an elementary observation: Individuals are altered and shaped through interaction with others. 
Equally basic is the observation that individuals can often decide with whom to interact. 
Both of these observations can be modeled by a single network, in which nodes correspond to individuals while links correspond to interpersonal connections. 
The dynamics of this network is governed by two processes: 
Topology-dependent transmission of dynamical states of individuals, and state-selective evolution of the links. 
Hence, the combination of the two gives rise to an adaptive network~\cite{GrossBlasius}.

Within the framework of contact processes on adaptive networks, attention has focused particularly on opinion formation \cite{EhrhardtMarsili, GilZanette, GrawbowskiKosinski, ZanetteGil, Bencziketal, HolmeNewman06, KozmaBarrat, KozmaBarrata, NardiniKozmaBarrat} and epidemic spreading \cite{GrossDommarBlasius, Zanette, GrossKevrekidis, RisauGusmanZanette, ShawSchwartz, ZanetteRisauGusman, Gross}.     

Comparing the models studied in the context of the different applications reveals many similarities and some distinct differences. 
Similarities are found mainly in the general set-up of the models.
First, the transmission of states is strictly limited to the neighborhood of a node. 
Second, to account for differences among individuals and to facilitate computation, the processes in the model are in general defined stochastically.
Third, concerning the topological evolution, the vast majority of models allows only for rewiring of links. 
In contrast to other processes, rewiring conserves the number of nodes and links, which is advantageous 
for numerical simulation.
Finally, all numerical models discussed in this chapter apply an asynchronous update procedure, in which a randomly selected node is updated in any one step.
This is believed to yield the best approximation to a continuous time system~\cite{DurrettLevin}. 

The differences between models of epidemics and models of opinion formation arise mainly from differences in the physics of the underlying real-world processes: 
In epidemics, there is an objective difference between infected and healthy individuals, and the processes are inherently state-specific: The infection can be transmitted along the links, while it is obvious that the same is not possible for the healthy state.
By contrast, in models of opinion formation the different opinions are in general treated equally and therefore appear symmetrically in the model.
One important consequence of the state-dependence of epidemic processes is that additional processes have to be introduced in the model if the number of states is increased. 
Indeed, many models of epidemics extend the scenario of healthy and infected individuals by additional states to model distinct temporal phases of the infection. 
If for instance a state is introduced, which corresponds to individuals that have recovered from the disease, new processes have to be formulated that govern transitions to and from this state. 
Conversely, the symmetry of state-dependence in opinion-formation processes enables us to increase the number of states without increasing the number of processes in the system. 
On the one hand this means that a system with a small number of opinions greater than two will behave very similarly to a system with just two opinions~\cite{NardiniKozmaBarrat}. 
On the other hand it allows to consider systems in which infinitely many opinions compete based on a finite number of processes.

Regardless of the model, the investigation of contact processes on adaptive networks poses characteristic difficulties. 
Full agent-based simulations are fundamentally inefficient.
In order to determine the long-term behavior of the system we have to simulate for a long time. 
During this time the simulation produces information, namely a dynamical trajectory, which comes at a computational cost although it is generally not used in the analysis of the system.  
By contrast, the theory of dynamical systems offers many tools, such as Newton's Method and bifurcation analysis, that enable us to determine the long-term behavior of the system directly.  
In order to apply these methods the adaptive network needs to be described in terms of emergent variables, governed by differential equations or discrete time maps.
For contact processes, convenient variables are the densities of certain subgraphs called network moments.
A \emph{moment expansion} of the dynamics results in an infinite cascade of differential equations. 
This cascade can be truncated by a \emph{moment-closure approximation} which is explained below.
In practice, it is often sufficient to approximate the network by a small number of differential equations (e.g.~3), which allows for analytical treatment of the system.

In this chapter, we aim to provide an overview of recent studies concerned with contact processes on adaptive networks. 
Throughout these studies certain system level phenomena, like the emergence of state homogeneous subpopulations, are found to recur. The underlying mechanisms of these phenomena are addressed and compared.
In Sec.~\ref{cpsec:2} various papers are reviewed which treat opinion formation by means of different models. A comparison of these models provides insights into the topic \emph{per se} and, moreover, into the relation between the microscopic rules and the system level behavior. 
Sec.~\ref{cpsec:3} focuses on models of epidemic spreading. The application of an moment closure approximation is demonstrated by means of the adaptive SIS-model. Thereafter we launch into a more general discussion of moment closures. 
In particular we emphasize that the adaptivity of the network improves the efficiency of this tool.
    
\section{Opinion formation - theme and variations}
\label{cpsec:2}
Models of opinion formation explore the spreading of opinions in social networks.
Current models assume that this spreading is governed by two competing processes: social adjustment and social segregation. 
The former means that connected individuals adjust their views, the latter that individuals maintain contacts preferentially to like-minded individuals. 
In general, both processes reduce the number of links between nodes with conflicting opinions and lead to the formation of homogeneous social communities holding a uniform opinion.
A network which is entirely composed of such \emph{consensus communities} is said to be in the \emph{consensus state}. 
While almost all models ultimately reach a consensus state, the convergence time $\tau_c$ and the distribution of community sizes $P_s$ can differ markedly depending on the relative rate of the competing processes. 

As interpersonal interactions are highly complex and difficult to capture in models, a variety of different modeling approaches have been proposed.
This diversity provides the opportunity to investigate which details of the microscopic description affect the system level properties. 
Below, models of opinion formation are compared that differ mainly in the three aspects subsequently described. 

The first aspect concerns the number of opinions in the model. 
As we have mentioned above, essentially two cases have to be distinguished: 
Models in which only two alternative opinions exist, and those in which individuals can choose from a continuous spectrum of opinions.
The first, so-called \emph{voter-like} approach models typical electoral decisions, where the number of choices is limited by the number of candidates. 
The second approach applies to opinions such as religious belief, where in principle an infinite number of choices exists. 

The second aspect in which models differ is the treatment of social segregation. 
A link that connects individuals with conflicting opinions can either be rewired or broken entirely.
In the first case the number of links is conserved, and therefore the process is reversible. 
In the second case the number of links is decreased, therefore the process is irreversible unless it is 
counteracted by another process in which new links are created. 
So far, the creation of links has hardly been considered in models of opinion formation as it causes numerical difficulties and introduces additional complexities. 
 
Another difference between the models is how the symmetry of social interactions is broken.
In almost all models of opinion formation adjustment of views is conceived as an asymmetric act.
However, in the absence of a parameter that measures the persuasive power or the social influence of an individual, the implementation of asymmetry between interacting nodes is arbitrary: If we first randomly chose a node $i$ and subsequently randomly chose one of its neighbors $j$, then $i$ might either adopt the opinion of $j$ or vice versa. The first option defines a so-called \emph{reverse}, the second a so-called \emph{direct update rule}. It is known that both rules result in qualitatively different behavior~\cite{NardiniKozmaBarrat}.

In the following we discuss four major contributions to the subject of opinion formation on adaptive networks. 
Section~\ref{subsec:contopi} focuses on a model by Holme and Newman, which features a continuous spectrum of opinions~\cite{HolmeNewman06}. Social segregation is modeled through rewiring and social adjustment through an reverse opinion update. 
The model which will be discussed in Sec.~\ref{subsec:two-valued choice} can be considered as opposite approach: In Ref.~\cite{GilZanette} Gil and Zanette investigate a voter-like model, in which social segregation is modeled through deletion of links. 
The model by Kozma and Barrat~\cite{KozmaBarrat}, which is discussed in Sec.~\ref{subsec:tolerance}, again considers the choice between infinitely many opinions. The main difference to Ref.~\cite{HolmeNewman06} is that social segregation and social adjustment are restricted by an additional parameter, which can be interpreted as bounded tolerance.
In Sec.~\ref{subsec:asymmetry} a paper of Nardini et.~al.~is addressed that compares two voter-like models, both of which use identical rewiring rules but differ with respect to the direct/reverse implementation of the asymmetric adjustment process~\cite{NardiniKozmaBarrat}.

\subsection{Continuous opinions}
\label{subsec:contopi}
Holme and Newman were the first to report that the diversity of opinions sustained in a society undergoes a phase transition if the relative rate of social adjustment and social segregation crosses a critical threshold~\cite{HolmeNewman06}. 
In their paper, they consider the case of opinions which are in principle unlimited in number. 
A node $n$ is initially assigned an opinion $g_n$ at random. 
In each timestep, a node $i$ is randomly chosen and updated in one of two ways: 
With probability $1-\phi$, $i$ is convinced by one of his neighbors $j$ and $g_i$ is set to equal $g_j$. 
With probability $\phi$, node $i$ randomly selects one of its links and reconnects it to a node with opinion $g_{i}$. 

Note that this parameterization in terms of $\phi$ is advantageous as only the relative rate of the two processes is important.
Rescaling the sum of the rates of the two processes to one normalizes the frequency of events to one per update and thus effects an optimization of simulation time. Although this is rarely spelled out this parameterization is indeed an event-driven simulation of the two competing processes following the Gillespie algorithm.

In simulations, the system ultimately approaches a consensus state, in which all individuals in the same connected component hold the same opinion. 
As mentioned above, there is no objective difference between different opinions. 
Thus, in analyzing the consensus state it is not of interest which particular opinions survive, but how many and how the followers are distributed. 
This information is captured by the component-size distribution $P_s$.  
 
\begin{figure}[h]
\sidecaption [t]
\includegraphics[width=0.6\textwidth]{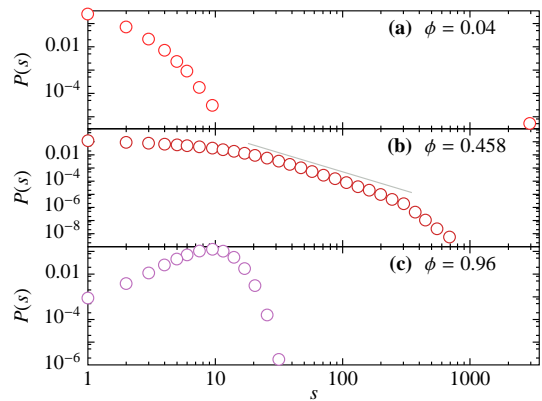}
\caption{Distribution of community sizes in the consensus state for $\phi$ below (a), at (b) and above the critical point (c). Numerical data are averaged over $10^4$ realizations for each value of $\phi$. $N=3200$, $\bar{k}=4$. Figure extracted from~\cite{HolmeNewman06}}
\label{cpfig:1}       
\end{figure}

Figure~\ref{cpfig:1} summarizes the dependence of $P_s$ on $\phi$. 
For $\phi=0$, no connections are rewired, so the component-size distribution of the initial random graph is conserved. In a random graph with mean degree $\langle k \rangle>1$ there is one giant component of the size 
$O(N)$ and $O(N)$ small components of size $O(1)$ (see Fig.~\ref{cpfig:1} a). We therefore find a large majority holding one opinion and many small groups holding different opinions.
For $\phi=1$, opinions never change, so the final cluster-size distribution equals the initial distribution of opinions. In particular the giant component splits into fragments of finite size (see Fig.~\ref{cpfig:1}(c)).

Applying a finite-size scaling analysis, Holme and Newman are able to show that a critical parameter value $\phi_c\approx0.458$ exists, at which a continuous phase transition takes place. 
At this transition the distribution of followers $P_s$ approaches a power-law (Fig.~\ref{cpfig:1}(b)). 

The convergence time $\tau_c$ needed to reach the consensus state is shown to scale differently in the regimes to both sides of the phase transition. 
For $\phi=1$ $\tau_c$ scales as $N$ and for $\phi=0$ as $\log(N)$. 
For $\phi\approx\phi_c$, $\tau_c$ obeys a scaling relation of the form $N^{-\gamma}$ with the critical exponent $\gamma=0.61\pm 0.15$ based on numerical simulations.

\subsection{Two-valued choice and irreversible discord} 
\label{subsec:two-valued choice}
The scenario that Gil and Zanette discuss in Ref.~\cite{GilZanette, ZanetteGil} deviates in two respects from the one investigated above.
Firstly, the regarded model is voter-like which means that choices are two-valued. 
Secondly, disagreeing neighbors break contact irrevocably. 
Starting from a fully connected network with randomly distributed opinions, conflicts are settled by convincing neighbors or cutting links. 
As above, rates of both processes are subsumed under one parameter $q$, which is defined as the probability of opinion transmission.  

The dependence of the community-size distribution on $q$ described by Gil and Zanette matches the results of Holme and Newman. 
Differences in the set-up are solely mirrored by ``boundary effects": 
In the absence of topological evolution the number of opinions in the final state equals the number of initially disconnected communities, which is one in the case under consideration and greater than one in \cite{HolmeNewman06}. 
In the opposite limit, i.e., without contact interactions, the number of disconnected communities in the final state equals the number of initial opinions, which is two in the model of Gil and Zanette and greater than two in that of Holme and Newman. 
For intermediate values of $q$ ($\phi$ respectively), the mean of the distribution $P_s$ shifts in both models from smaller to larger $s$ as contact interactions gain influence.

Let us now discuss the underlying mechanisms that lead to the formation of similar community-size distributions in the two different models. 
In both models, the processes of social adjustment and social segregation occur only on links between disagreeing neighbors, which we therefore call \emph{active links}.
The consensus state is reached when all of these active links have vanished. 
Although segregation is modeled by rewiring in~\cite{HolmeNewman06} and by cutting links in~\cite{GilZanette} the effect is in both cases a reduction of active links. 
Social adjustment results in both models either in an activation or a deactivation of links. 
Note however, that in voter-like models adjustment reverses the state of all links connecting to the target node. By contrast, in models with continuous opinions, active links connecting to the target node may remain active. 
Nevertheless, we know that both models eventually reach consensus even without segregation, therefore social adjustment has to decrease the number of active links \emph{in average}. 

While the effect of both, adjustment and segregation, is in the long run a reduction of active links, both processes have a different impact on the consensus time $\tau_c$. 
As we have seen above, consensus through social adjustment requires a convergence time which scales like $N$.  
Social segregation significantly accelerates consensus but separates neighbors, whose opinions could in the long term have converged through social adjustment. 
Thus, increased segregation leads to increased fragmentation, which explains the segregation-rate dependent changes of the distribution $P_s$ as well as their independence of the differences between \cite{HolmeNewman06} and \cite{GilZanette}.  

\begin{figure}[h]
\sidecaption
\includegraphics[width=0.6\textwidth]{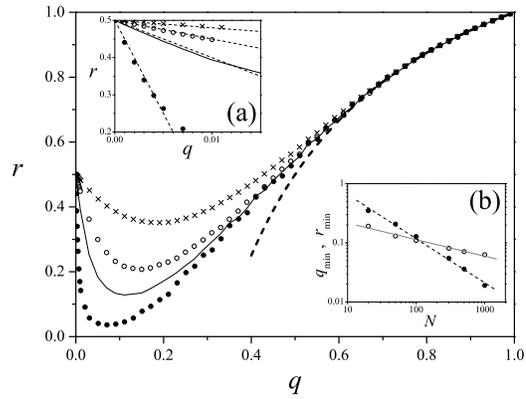}
\caption{Fraction $r$ of remaining links as a function of the parameter $q$. Different symbols correspond to different system sizes, $N=20$ $(\times)$, $50$ $(\circ)$, $100$ $(-)$ and $500$ $(\bullet)$. The dashed line represents the analytical approximation for large $N$. Insert (a): detailed view of the same data for small $q$. Insert (b): Position $q_{min}$ $(\circ)$ and depth $r_{min}$ $(\bullet)$ of the minimum of $r$ as a function of $N$. Figure extracted from ~\cite{GilZanette}}
\label{cpfig:2}       
\end{figure}

The link-deletion process in the model of Gil and Zanette reveals a phenomenon, which is not obvious in the model of Holme and Newman.
Even though the number $\tau_c(q)$ of events necessary to reach consensus decreases with decreasing $q$, the number of segregation events $(1-q)\tau_c(q)$ depends non-monotonically on $q$.
As shown in Fig.~\ref{cpfig:2}, a critical parameter value $q_{min}$ exists, at which the fraction $r$ of remaining links in the consensus state is minimized, i.~e. at which a maximum number of deletion events occur. 
This can be understood intuitively:
The fraction of remaining links $r$ is minimized if between two subsequent opinion flips the majority of active links is deleted but no consensus communities are isolated.  
In such a situation an opinion flip almost exclusively activates links, the majority of which will in turn be deleted.
If less than the critical number $1-q_{min}$ of active links are deleted, the opinion flip not only activates but also inactivates links. These inactive links, unless reactivated later, are not available for deletion, and thus $r$ increases.
If on the other hand more than $1-q_{min}$ active links are deleted, the probability increases that consensus communities are isolated. Internal links of such communities can not be activated in subsequent adjustment events, increasing $r$. 
   
Based on the similarity of the two compared models, it is arguable whether the critical parameter $\phi_c$ in ~\cite{HolmeNewman06} corresponds to the same phase transition as $q_{min}$ in ~\cite{GilZanette}. 
Encouraging in this regard are recent findings of Vazquez et al. that indicate the existence of a generic fragmentation transition for different voter-like models \cite{VazquezEguiluzSanMiguel}. 
One may argue that the critical parameter $\phi_c$ is independent of the system size while $q_{min}$ decreases with growing $N$ (cf. Fig.~\ref{cpfig:2}(b)). 
However, the $N$-dependence of $q_{min}$ is only a result of the initial conditions chosen in~\cite{GilZanette}: 
As the initial graph is fully connected, an opinion-flip event affects $O(N)$ links, whereas a link-deletion event affects one link regardless of the system size. 
The relative rate of adjustment and segregation events, which minimizes the fraction of remaining links, therefore approaches zero if $N$ goes to infinity.

\subsection{The influence of bounded tolerance}
\label{subsec:tolerance}
The influence of tolerance on opinion formation is investigated in~\cite{KozmaBarrat, KozmaBarrata}. 
In these papers, Kozma and Barrat consider a scenario where opinions can take continuous values. 
A global parameter $d$ is introduced describing the tolerance range of individuals. 
If opinions of neighbors are closer than the tolerance range, i.e.,~if $\left|g_i-g_j\right|<d$, both adopt the mean opinion with probability $1-w$. 
If opinions of neighbors differ more than the tolerance range, $i$ rewires with probability $w$ to a randomly chosen node $k$. 

\begin{figure}[h]
\sidecaption
\vspace{0.45cm}
\includegraphics[width=0.55\textwidth]{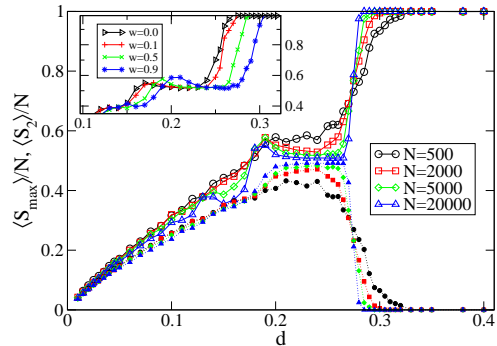}
\caption{Size of the largest (open symbols) and second largest (filled symbols) homogeneous opinion cluster as a function of the tolerance $d$. The color coding for the system size is the same for the largest and second largest cluster. Inset: Size of the largest opinion cluster as a function of $d$ for different rewiring rates $w$. Figure extracted from~\cite{KozmaBarrat}.}
\label{cpfig:3}       
\end{figure}

If defined in this way, bounded tolerance has two different effects: 
On the one hand it reduces the selectivity of social segregation.
On the other hand it enhances selectivity of social adjustment. 
To illustrate these points let us first consider the effect of bounded tolerance in the absence of segregation.
In this case a consensus opinion in a component is only reached if tolerance intervals of neighbors overlap. 
Otherwise ``tolerance patches" may form in which nodes are locally in consensus but do not communicate with nodes outside the patch. In these tolerance patches conflicting opinions can survive indefinitely and
thus the equivalence of topological components and consensus communities in the final state is broken.
However, to describe the final state we stick with the terminology, which was introduced above, and only adapt the meaning of the term ``consensus community" slightly: 
Used in the present context, it refers to communities of like-minded individuals that are necessarily connected among themselves but not necessarily isolated from individuals of other communities. 
Kozma and Barrat show that, in the absence of segregation, three parameter regimes can be identified (cf. Fig.~\ref{cpfig:3}): 
For large tolerance $d$, the set-up matches the $\phi=0$ case in Ref.~\cite{HolmeNewman06}. 
Consequently, the system reaches a state where nearly all nodes are belong to a single community of like-minded individuals. 
Only when $d$ falls below a critical value $d_c\approx 0.256$, the enhanced selectivity of social adjustment is noticeable. 
Then, the final state becomes polarized, i.~e.~two macroscopic communities are observed to coexist with a number of finite size communities. 
Finally, for very small $d$, an extensive number of small communities form an fragmented final state.
   
The onset of rewiring is found to have different effects in the different parameter regimes. 
On the one hand, it impedes complete consensus: the larger the rewiring rate, the larger tolerance values are necessary to reach complete consensus (cf. inset Fig.~\ref{cpfig:3}). 
On the other hand, in the fragmented regime, it leads to an enlargement of the consensus-community sizes. 
This can be explained as follows:   
For large tolerance, neighbors with overlapping tolerance intervals prevail.
Opinions of neighbors that differ more than $d$ are altered through interaction with other neighboring nodes and eventually become closer than $d$. 
Hence, the key to the formation of extended communities lies in the possibility of repeated contact interactions. 
As in the previously studied models, rewiring disconnects communities prematurely and thereby impedes complete consensus. 
For small tolerance, the limiting factor for the size of consensus communities is the small number of neighbors with overlapping tolerance intervals. In this situation, rewiring allows each node to find those nodes it can communicate with and thus facilitates the merging of small groups. 

Indeed, the two different effects of rewiring can also be seen in the model of Holme and Newman. 
The initial giant component is split due to rewiring.
The initial components of finite size, which corresponds to the limit of small tolerance, gain size (cf. Fig.~\ref{cpfig:1}).   

\subsection{Asymmetric insertion of influence}
\label{subsec:asymmetry}
All models presented so far feature asymmetric interactions between a randomly chosen node and a random neighbor. 
In contrast to a randomly chosen node a random neighbor is not drawn in an unbiased way -- it is reached by following a link and therefore nodes with higher degree are preferentially selected as random neighbors.
The symmetry of node and neighbor in the rules of the model is in some cases broken by definition of the contact process ~\cite{GilZanette, HolmeNewman06}, and in others by the definition of the rewiring mechanism~\cite{HolmeNewman06, KozmaBarrat}. The effect of the asymmetry of the interactions is studied by Nardini et al. \cite{NardiniKozmaBarrat} via a mean field analysis. 
Nardini et al.~show that, in case of inhomogeneous networks, the implementation of the asymmetry may decisively influence the behavior of the system. 
They compare two voter-like models that differ with respect to the asymmetry of the opinion updates. 
In both models each timestep begins with choosing an individual $i$ and one of its neighbors $j$ at random. 
If $i$ disagrees with $j$, it cuts the link and establishes a new link to a randomly chosen node $k$ with probability $\phi$. With probability $1-\phi$, one of the two convinces the other of its opinion. 
The difference in the models lies in the node that is convinced. The first alternative is a reverse voter-like model (rVM), in which $i$ is convinced by $j$. The second alternative is the direct voter-like model (dVM), in which $j$ is convinced by $i$. 

Simulations show that in both models nodes of the majority opinion have a higher average degree than nodes of the minority opinion. 
Nodes with high degree, however, are preferentially selected as random neighbors $j$ \cite{AlbertBarabasi, Newman}, and 
hence, the random neighbor $j$ is likely to hold the majority opinion. 
That is, of two dissenting neighbors, a random node $i$ and its random neighbor $j$, $i$ probably holds the minority and $j$ the majority opinion.
In the rVM the majority opinion reproduces itself as $j$ convinces $i$. Thus, once a disparity between both opinions emerges it increases. 
By contrast, in the dVM the majority opinion is repressed as $j$ is convinced by $i$. Any disparity in the opinion distribution will therefore undergo damping.            

\begin{figure}[h]
\begin{minipage}[t]{0.465\textwidth}\raggedleft
\includegraphics[width=\textwidth]{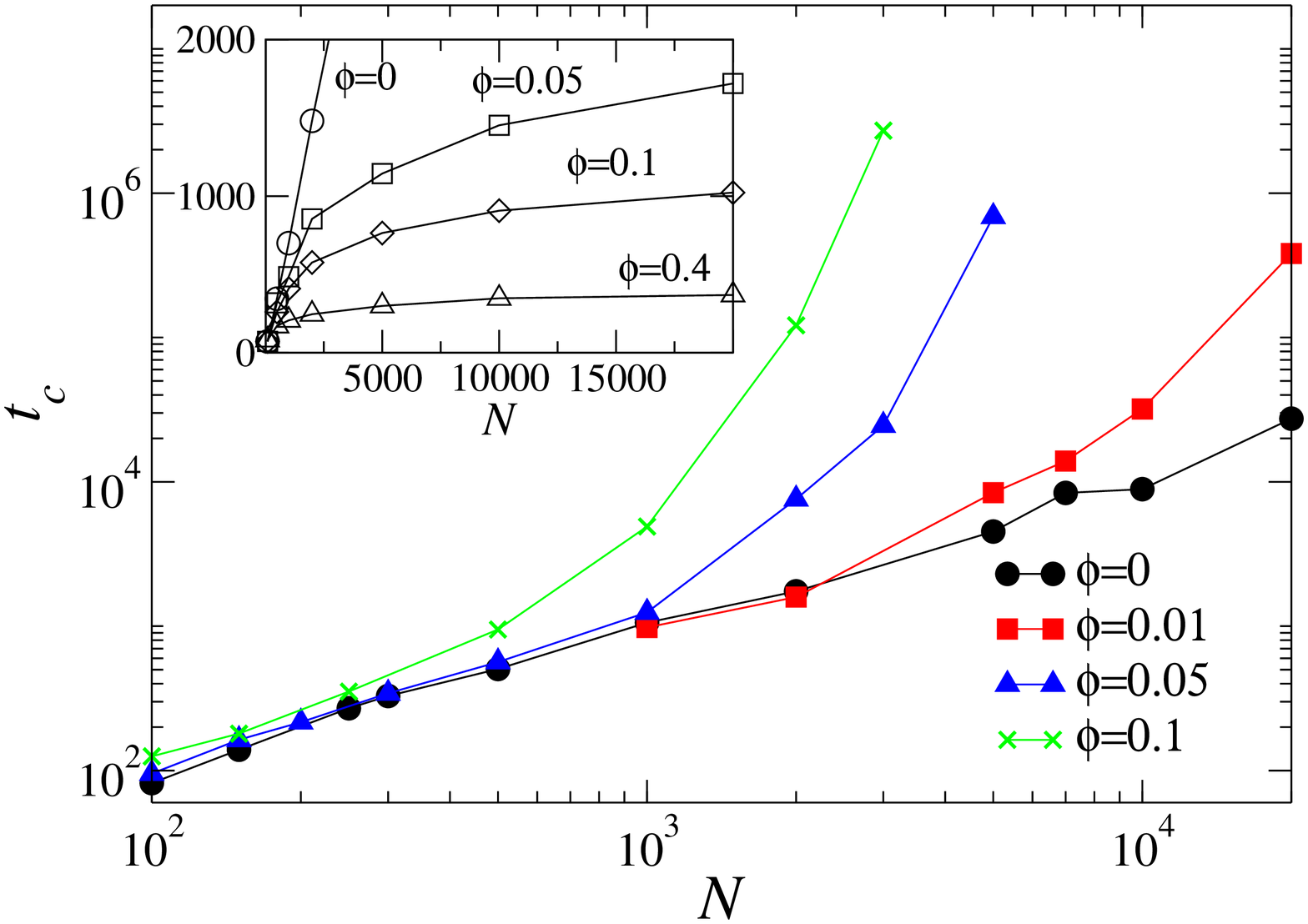}
\end{minipage}\hspace{0.05\textwidth}
\begin{minipage}[t]{0.473\textwidth}
\includegraphics[width=\textwidth]{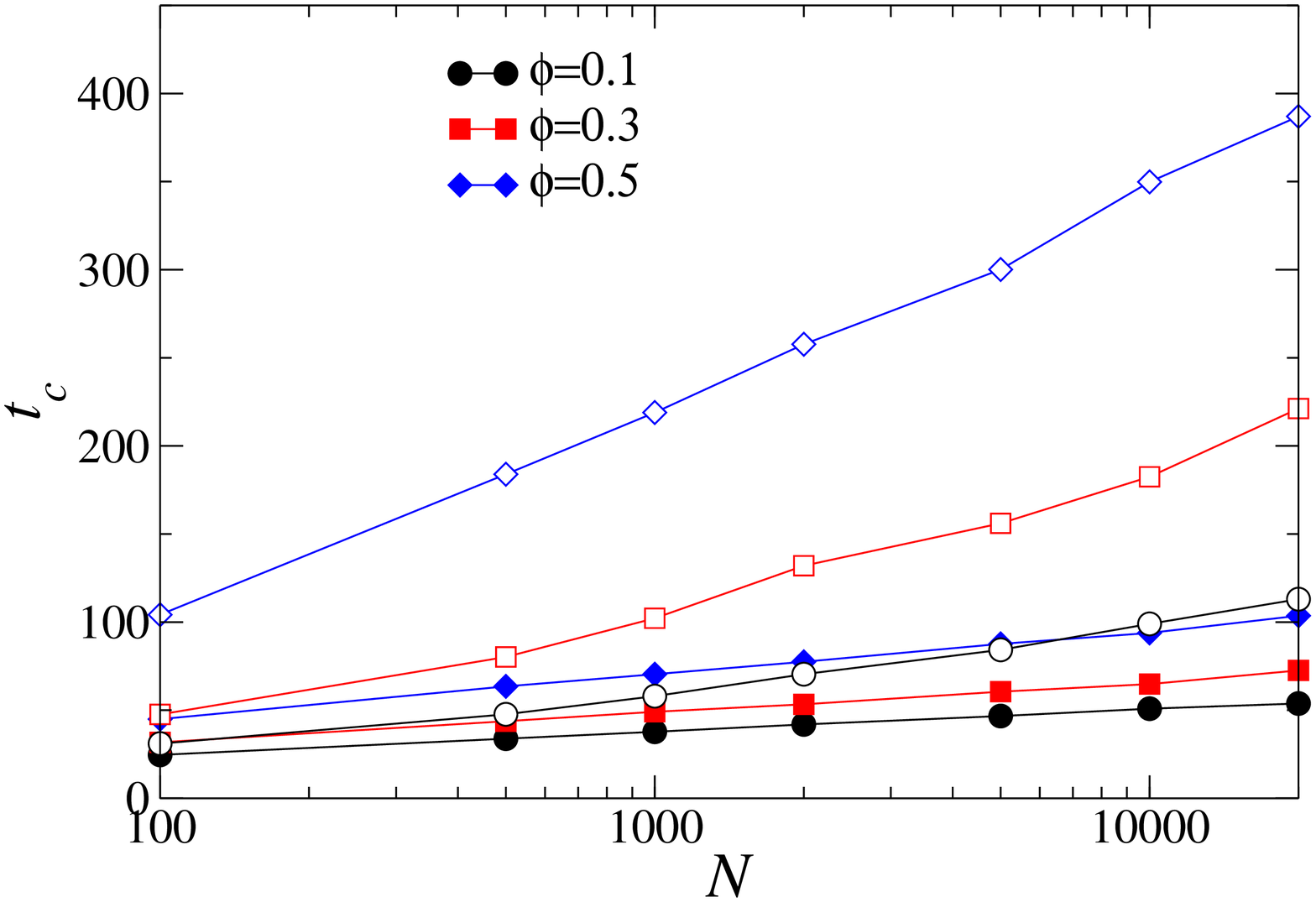}
\end{minipage}
\begin{picture}(200,1)
\put(15,1) {(a)}
\put(190,1){(b)}
\end{picture}
\caption{(a) Convergence time for the reverse voter-like model as a function of the system size for various rewiring rates. Inset: same for the direct voter-like model. (b) Convergence time for the direct (filled symbols) and reverse naming game. For each parameter set, data are averaged over 100 realizations of the system. Figure extracted from ~\cite{NardiniKozmaBarrat}.}
\label{cpfig:4}       
\end{figure}

In summary, adaptivity generates a positive feedback in case of the rVM impelling the system toward an accelerated consensus. 
In case of the dVM the generated feedback is negative resulting in a dynamical state where both opinions are in average equally represented.
For the parameter values chosen in the paper no consensus is reached in the latter case. Nevertheless, small networks fluctuations may still take the system eventually to an absorbing state in which one opinion vanishes.  
The different routes to consensus are reflected in the specific convergence time $\tau_c(N)$ observed in numerical simulations (see Fig.~\ref{cpfig:4}). 
For the rVM, $\tau_c$ displays a logarithmic scaling behavior $\tau_c(N)\propto\ln(N)$ while for the dVM, $\tau_c(N)$ grows exponentially with the system size.    

Remarkably, the qualitative differences between the dVM and the rVM are settled if an additional neutral state is introduced, which is the case in the so-called naming game. 
In this scenario, change of opinion is impeded in the sense that individuals have to pass a transient state before defecting to the opposite view. As long as an individual is in this state, it is accessible for convincing attempts from representatives of both opinions. 
To model the naming game Nardini et al.~choose the following implementation: The competing opinions are assigned with the values $+1$ and $-1$ and the additional neutral state is denoted by $0$. Contact interactions between disagreeing neighbors alter the state of the passive node by $\pm1$, whereby the sign depends on the state of the active node. If the active individual is in the neutral state, it chooses to represent one of the opinions $+1$ or $-1$ at random. In analogy to the direct and reverse voter-like models, a distinction can be made between the direct and reverse naming game depending on whether the random node or the random neighbor takes the active part. 

On a static network the consensus time $\tau_c$ in naming games is known to scale like $\ln(N)$. 
Simulations yield that in the adaptive case $\tau_c(N)$ remains logarithmic irrespective of the chosen modality of asymmetric opinion update (cf. figure ~\ref{cpfig:4}(b)). 
This deviation from the behavior of voter-like models is elegantly explained by Nardini et al.: 
Interactions via links between followers of the competing opinions comply with the dynamics of the dVM (rVM respectively) and exhibit the characteristic negative (resp. positive) dynamical feedback. 
However, interactions with neutral nodes exert a positive feedback regardless of whether the direct or reverse rule is used. As links to neutral nodes are far more common than links to nodes of the opposing opinion, these links dominate the behavior and hence in total a positive feedback is observed.

\subsection{Other approaches}
A slightly different setup for the contact process is explored by Benczik et al. \cite{Bencziketal}, Grabowski and Kosi\'nski \cite{GrawbowskiKosinski} and Erhardt et al. \cite{EhrhardtMarsili}. 
Instead of occasional interactions between two randomly chosen neighbors, they consider situations where a node is updated by evaluation of all influences from its entire neighborhood. 
All three models capture various additional properties. 
Thus, besides internal state dynamics Ehrhardt et al.~include adjustable link creation and removal processes as well as sophisticated partner selection mechanisms. 
Equally elaborated are the topological evolution rules Grabowski and Kosi\'nski use: the idea of bounded tolerance is combined with a set of parameters that model individually distinct sociability. Furthermore, some links, which represent basic connections like family ties, are excluded from the topological changes.
Finally, Benczik et al. investigate a topological evolution rule in which a continuous parameter captures the individuals' tendency to rather avoid or seek contact with dissenting individuals.  
For more details we refer to the original publications.   

An interesting enhancement of the concept of bounded tolerance is discussed in~\cite{GargiuloMazzoni}.
In this paper, Gargiulo and Mazzoni replace the global tolerance parameter $d$ by state dependent tolerance. 
The underlying hypothesis is that in realistic systems tolerance decreases if opinions get extreme. 
Regrettably, the approach is so far only explored in simulations in which tolerance-dependent segregation and tolerance-dependent adjustment occur in consecutive temporal phases of the evolution. 
So, an exploration in the context of adaptive networks remains to be done.

\section{Epidemic spreading and Moment closure}
\label{cpsec:3}
Subsequently, we focus on models of epidemic spreading, the second intensively studied topic in the class of contact processes. 
Though both, models of epidemic spreading and models of opinion formation, base on the concept of locally transmitted properties, epidemiological models are essentially distinguished from those discussed in Sec.~\ref{cpsec:2}.
In epidemiological models, different single node states signify different stages of a disease.
This interpretation imposes far-reaching restrictions on the processes modeling the transitions between the states.
Construing states as stages of a disease directly attaches a meaning to the transitions between states
Thus, transitions can only occur between appointed states, which reminds of the naming game but differs from the scenario reviewed in Sec.~\ref{subsec:contopi}.
Moreover, the asymmetry of the contact process is determined by the qualitative differences of the states: Via a link between an infected and a healthy individual, only the infected state can be spread. 
This is contrary to models of opinion formation where the asymmetry of the convincing act was implemented arbitrarily.    
Hence, for each state in an epidemiological model we have to formulate specific processes that govern transitions to and from this state. 
In practice, the transmission of the disease is the only real contact process, while all other processes describe the subsequent progession through epidemic stages which happens only locally.

While models of opinion formation were shown to vary with respect to the implementation of asymmetry, the number of states, and the topological evolution rules, models of epidemic spreading do not exhibit any variations with respect to the implementation of asymmetry.
Variations in the number of single-node states are impeded as the introduction of new states necessitates the introduction of new processes.
Variations of the topological evolution rules are discussed, however to a minor extend.   

The substantial coherence among different adaptive-network models of epidemiological processes allows us to focus exemplarily on the adaptive SIS-model, which features only two states called S, for susceptible, and I, for infected.
By means of this simple model, we illustrate the conceptual and methodical framework likewise applying for more complicated scenarios (Sec.~\ref{subsec:SIS}).
In particular, we demonstrate the use and handling of \emph{moment-closure approximations}, a common tool in epidemiology \cite{KeelingRandMorris,ParhamSinghFerguson,PeyrardDieckmannFranc08}.
Section~\ref{subsec:Extensions} launches variations and extensions of the basic SIS model, that aim for more realism \cite{Zanette,ZanetteRisauGusman,ShawSchwartz}. 

\subsection{The adaptive SIS model}
\label{subsec:SIS}
 
The simplest model, in which epidemic dynamics and topological evolution can be combined is the SIS model.  
It describes a scenario in which each individual within a social network is either susceptible (S) to the disease under consideration or infected (I).
 Contacts between individuals are denoted as SS-links, SI-links, and II-links according to the states of the individuals they connect. 
Susceptible individuals can become infected if they are in contact with an infected individual. 
The transmission of the disease along a given SI-link is assumed to occur at a rate $p$. 
Once an individual has been infected she has a chance to recover, which happens at a rate $r$ and immediately returns the individual to the susceptible state. 
In the adaptive SIS model proposed by Gross et al. \cite{GrossDommarBlasius} another process completes the circle of infection and recovery: 
If a susceptible individual is connected to an infected individual she may want to break the link and instead establish a new link to another susceptible. On a given SI-link this rewiring occurs at a rate rate $w$.

Note that the rewiring process has been introduced `optimistically': Only susceptible nodes rewire, and they manage unerringly to rewire to a node that is also susceptible. Under these conditions rewiring always reduces the number of links that are accessible for epidemic spreading and therefore the \emph{prevalence} of the disease, i.e., the density of infected, is always reduced by this form of rewiring behavior. 
Less optimistic rewiring rules have been explored by Zanette \cite{Zanette}, and Zanette and Risau-Gusm\'an \cite{RisauGusmanZanette, ZanetteRisauGusman} and will be addressed in Sec.~\ref{subsec:Extensions}.     

Let us now study the dynamics of the adaptive SIS model with the tools of nonlinear dynamics. 
For this purpose we need to derive a low-dimensional emergent-level description of the system. 
Convenient observables, so-called moments, are given through the densities of certain subgraphs. 
The number of links contained in such a subgraph is called the order of the respective moment. 
Dynamical properties of the moments, averaged over many realizations of the stochastic process, can be summarized in a system of ODEs. 
Due to the contact process, however, dynamics of moments of order $n$ essentially depend on moments of order $n+1$, resulting in an infinite cascade of differential equations. 
Its truncation necessitates an approximation of higher order moments in terms of lower order moments, the so-called moment closure approximation.

Below, we will derive an emergent-level description of the adaptive SIS model using moment closure approximation.  
In the SIS model, the moments of zeroth order are the densities of infected and susceptibles, $[I]$ and $[S]$. 
First order moments are the per-capita densities of SS-, SI- and II-links, $[SS]$, $[SI]$ and $[II]$, and second order moments the densities of triplets $[ABC]$ with a given sequence of states $A,B,C\in\left\{I,S\right\}$. 
Due to the conservation relations $S+I=1$ and $[SS]+[SI]+[II]=\langle k \rangle$ the dynamics of the zeroth and first order moments are entirely captured by the balance equations for $[I]$, $[SS]$, and $[II]$. 
A further advantage of the normalization relations is that we can write all subsequent equations as if we were dealing with a number of individual nodes and links instead of densities.  

Let us start by writing a balance equation for the density of infected nodes. Infection events occur at the rate $p[SI]$ increasing the number of infected nodes by one; Recovery events occur at a rate $r[I]$ and reduce the number of infected nodes by one. This leads to 
\begin{equation}
\label{eqInfected}
\frac{\rm d}{\rm dt} [I]=p[SI]-r[I].
\end{equation}
The equation contains the (presently unknown) variable $[SI]$ and therefore does not yet constitute a closed model. One way to close the model were a mean field approximation, in which the density of SI-Links is approximated by $[SI] \approx \langle k \rangle [S][I]$. However, in the present case this procedure is not feasible: Rewiring does not alter the number of infected and hence does not show up in Eq.~(\ref{eqInfected}). Thus the mean-field approximation is not able to capture the effect of rewiring. Instead, we will treat $[SI]$, $[SS]$, and $[II]$ as dynamical variables and capture their dynamics by additional balance equations. This approach is often called moment expansion as the link densities can be thought of as the first moments of the network.

As stated above, it suffices to derive balance equations for the densities of SS- and II-links. The density of SI-links can then be obtained from the conservation relation. First the II-links: A recovery event can destroys II-links if the recovering node was part of such links. The expected number of II-links in which a given infected node is involved is $2[II]/[I]$. (Here, the two appears since a single II-link connects to two infected nodes.) Taking the rate of recovery events into account, the total rate at which II-links are destroyed is simply $2r[II]$. 

To derive the rate at which II-links are created is only slightly more involved. In an infection event the infection spreads across a link, converting the respective link into an II-link. Therefore every infection event will create at least one II-link. However, additional II-links may be created if the newly infected node has other infected neighbors in addition to the infecting node. In this case the newly infected node was previously the susceptible node in one or more ISI-triplets. Thus, we can write the number of II-links that are created in an infection event as $1+[ISI]/[SI]$. In this expression the `1' represents the link over which the infection spreads while the second term counts the number of ISI-triplets that run through this link. Given this relation we can write the total rate at which II-links are created as $p[SI](1+[ISI]/[SI])=p([SI]+[ISI])$. 

Now the SS-links: Following a similar reasoning as above we find that infection destroys SS-links at the rate $p[SSI]$. Likewise SS-links are created by recovery at the rate $r[SI]$. In addition SS-links can also be created by rewiring of SI-links. Since rewiring events occur at a rate $w[SI]$ and every rewiring event gives rise to exactly one SS-link the total rate at which rewiring creates SS-links is simply $w[SI]$.

Summing all the terms, the dynamics of the first moments can be described by the balance equations
\begin{eqnarray} \label{eqNotClosedA}
\frac{\rm d}{\rm dt} [SS] &=& (r+w)[SI] - p [SSI] \\\label{eqNotClosedB}
\frac{\rm d}{\rm dt} [II] &=& p([SI]+[ISI]) - 2r[II].  
\end{eqnarray} 
Again, these equations do not yet constitute a closed model, but depend on the unknown second moments $[SSI]$ and $[ISI]$. However, the first order-moment expansion captures the effect of rewiring. While we will return to the equation above later, a feasible way of closing the system is to approximate the second moments by a mean-field-like approximation: the \emph{pair approximation}. 

Let us start by approximating $[ISI]$. One half of the ISI-triplet is actually an SI-link, which we know occurs at the density $[SI]$. In order to approximate the number ISI-triplets running through a given link we have to calculate the expectation value of the number of \emph{additional} infected nodes that are connected to the susceptible node. For this purpose let us assume that the susceptible node of the given SI-link has an expected number of $\langle q \rangle$ links in addition to the one that is already occupied in the SI-link. Every one of these links is an SI-link with probability $[SI]/(\langle k \rangle S)$. (Here, we have neglected the fact that we have already used up one of the total number of SI-links. This assumption is good if the number of SI-links is reasonably large.) Taking the density of SI-links and the probability that they connect to additional SI-link into account we obtain 
\begin{equation}
\label{eqClosure}
[ISI]= \kappa \frac{[SI]^2}{[S]}  
\end{equation}              
where $\kappa=\langle q \rangle / \langle k \rangle$ remains to be determined. The quantity $\langle q \rangle$ that appears in $\kappa$ is the so-called \emph{mean excess degree}. Precisely speaking it denotes the expected number of additional links that are found by following a random link. 

Subsequently we will assume that $\kappa=1$. This assumption is substantiated in the reasoning of Ref.~\cite{Gross}. Here, we only state that it allows us to approximate the density of triplets by $[ISI]=[SI]^2/S$, and following a similar argumentation $[SSI]=2[SS][SI]/[S]$. Substituting these relations into the balance equations we obtain a closed system of differential equations

\begin{eqnarray}
\frac{\rm d}{\rm dt} [I]&=&p[SI]-r[I]\label{eqMotionA}\\
\frac{\rm d}{\rm dt} [SS] &=& (r+w)[SI] - 2p[SI] \frac{[SS]}{[S]} \\
\frac{\rm d}{\rm dt} [II] &=& p[SI](1+\frac{[SI]}{[S]}) - 2r[II]\ . \label{eqMotionC}
\end{eqnarray}
  
The system of differential equations can now be studied with the tools of dynamical systems theory.
Gross et al. compare the analytical results thus obtained with detailed-level simulations of the full model and find both in very good agreement~\cite{GrossDommarBlasius}. 
This indicates a high accuracy of the emergent-level description~(\ref{eqMotionA}--\ref{eqMotionC}). 

\begin{figure}[h]
\sidecaption[t]
\includegraphics[width=0.6\textwidth]{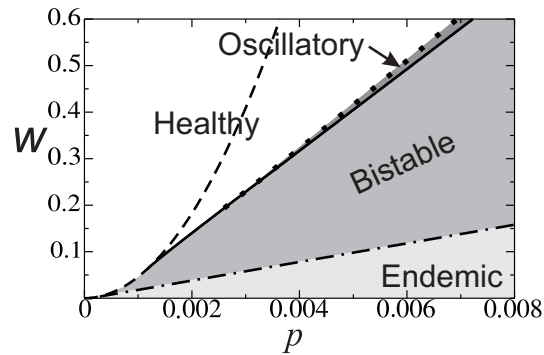}
\caption{Two parameter bifurcation diagram showing the dependence on the rewiring rate $w$ and the infection probability $p$ at fixed recovery rate $r=0.002$. Figure extracted from \cite{GrossDommarBlasius}.}
\label{cpfig:5}       
\vspace{-0.5cm}
\end{figure}

In contrast to the models of opinion formation, which have been discussed in Sec.~\ref{cpsec:2}, the adaptive SIS model features three instead of two processes. Therefore the dynamics in the SIS-model depends on two 
free parameters. Figure~\ref{cpfig:5} shows the two parameter bifurcation diagram which results from the analysis of Eqs.~(\ref{eqMotionA}--\ref{eqMotionC}).    
In the white and light gray regions there is only a single attractor, which is a healthy state in the white region and an endemic state
in the light gray region. In the medium gray region both of these states are stable. Another smaller region of bistability is shown in
dark gray. Here, a stable healthy state coexists with a stable epidemic cycle.
The transition lines between these regions correspond to saddle-node (dashed), Hopf (continuous), and cycle fold (dotted) bifurcations.
The dash-dotted line marks a transcritical bifurcation that corresponds to the threshold at which epidemics can invade the disease free system. 

Although the SIS-model, at first glance, differs strongly from the models of opinion formation, some interesting parallels appear. As in the models of opinion formation high rewiring rates break the network into  ``consensus'' communities in which all nodes are either susceptible or infected. In this way the contact process, infection, is impeded. This is for instance reflected in a strong increase of the invasion threshold (dash-dotted) with increasing rewiring rate. However, in contrast to the models of opinion formation, the dynamics does not freeze in the consensus state as recovery can still take place.

Another feature of the epidemic model that is not observed in the models of opinion formation is the bistable region. In this region an established epidemic can survive at high prevalence, while epidemics cannot invade a disease-free network. This region appears since the disease suppressing effect of segregation becomes weaker at high prevalence: In contrast to opinion formation models in which both opinions are treated equally, a small community of infected can be more easily isolated than a small community of susceptibles. This asymmetry arises as the links are always rewired into the susceptible community, which is irrelevant if the susceptibles are in the majority, but leads to a sharp rise in the the connectivity of susceptibles if they are in the minority.
However, high connectivity of susceptibles speeds up the infection process which competes with segregation. 
Under certain conditions the competition of the two effects can lead to oscillatory dynamics. Both the appearence of bistability and oscillations can therefore be linked directly to the explicit asymmetry that is introduced in the epidemic model.

\subsection{Other approaches}
\label{subsec:Extensions}
Variants of the adaptive SIS model, in which not only susceptible but also infected individuals may rewire their links, have been explored by Zanette and Risau-Gusm\'an \cite{RisauGusmanZanette, ZanetteRisauGusman}. 
In these works, the authors prove that rewiring remains advantageous for suppressing the disease even if the isolation of infected agents is modeled to be less effective than in \cite{GrossDommarBlasius}.

In \cite{GrossKevrekidis} Gross and Kevrekidis consider an adaptive SIS model in which the effectivity of 
of rewiring increases with increasing prevalence of the disease. In this case oscillations can be observed in a much larger parameter range and with significantly increased amplitude. 

Models with additional epidemic states have been studied by Shaw and Schwartz \cite{ShawSchwartz} and Risau-Gusm\'an and Zanette \cite{RisauGusmanZanette}. Moreover, Shaw and Schwartz also investigate 
the effect of noise on the system. This work is reviewed in the subsequent chapter.  

\section{Summary and Outlook}

In this chapter, we have reviewed a selection of recent papers concerned with opinion formation and epidemic spreading on adaptive networks.  

Comparing the reviewed approaches, we have focused on three major aspects in which models differ: First the number of single-node states a model captures, second the topological evolution rules, and third the way in which the symmetry of interactions is broken. 
In models of opinion formation, differences in the most subtle aspect, namely the direct or reverse implementation of the opinion update, have crucial impact on the system's behavior. By contrast differences in the two other aspects lead only to minor changes. 

In models of epidemic spreading, the asymmetry of interaction is inherent in the modeled situation and can therefore not be modified. We have argued that this intrinsic asymmetry is directly linked to the appearence of bistability and oscillations observed in the epidemic model.  
  
In all reviewed models, rewiring or cutting links lead to the formation of state-homo\-geneous subpopulations, providing an example for the appearance of global structure from local rules. The subpopulations exhibit different degree distributions if the rewiring rule is sensitive to differences between states, either externally imposed as in the epidemic model, or self-organized as in the models of opinion formation. 

The coupling of state-specific degree distributions and asymmetric exertion of influence can stabilize the system in a state in which two states survive at finite density. This can be observed both in the direct voter 
model and in the adaptive SIS model. In the latter case the dynamics go on indefinitely as no absorbing state can be reached at finite density of infected because of the local processes, i.e.~recovery.   

A central theme of this book, which appears clearly in this chapter, is that in the investigation of adaptive networks common themes are frequently found in models from very different backgrounds. This shows that adaptive networks, which have emerged from many different disciplines almost at the same time, start to grow together. Certainly more investigations are necessary, but the goal of a unifying theory of adaptive networks, is slowly emerging. In the future steps toward this goal, analytical approximations such as the moment closure approximation described here, will be of central importance, as they allow for a rigorous mathematical treatment and generalization of the observed phenomena. 


\end{document}